\documentstyle[12pt]{article}
\def\grad{\vec{\nabla}}
\begin{document}
\begin{center}
{\bf Speculations on primordial magnetic helicity}\\[.3in]
John M. Cornwall*\\[.2in]
{\it Department of Physics and Astronomy, University of California\\
405 S. Hilgard Ave., Los Angeles Ca 90095-1547}\\[.2in]
{\bf Abstract}
\end{center}
We speculate that above or just below the electroweak phase transition
magnetic fields are generated which have a net helicity (otherwise said, a
Chern-Simons term) of order of magnitude $N_B + N_L$, where $N_{B,L}$ is the baryon or
lepton number today.  (To be more precise requires much more knowledge of B,L-generating mechanisms than we currently have.)  Electromagnetic helicity generation is associated (indirectly)
with the generation of electroweak Chern-Simons number through B+L
anomalies.  This helicity, which in the early universe is some 30 orders of magnitude greater than what would be expected from fluctuations alone in the absence of B+L violation, should be reasonably well-conserved through the
evolution of the universe to around the times of matter dominance and
decoupling, because the early universe is an excellent conductor.  
Possible consequences include early structure formation;
macroscopic manifestations of CP violation in the cosmic magnetic field
(measurable at least in principle, if not in practice); and an inverse-cascade
dynamo mechanism in which magnetic fields and helicity are unstable to transfer
to larger and larger spatial scales.  We give a quasi-linear treatment of the general-relativistic MHD inverse cascade instability, finding substantial growth for helicity of the assumed magnitude out to scales $\sim l_M\epsilon^{-1}$, where
$\epsilon$ is roughly the B+L to photon ratio and $l_M$ is the magnetic correlation length.  We also elaborate further on an earlier
proposal of the author for generation of magnetic fields above the EW phase transition.\\[.6in]
UCLA/97/TEP/7 \mbox{} \hfill March 1997\\
\footnoterule
\noindent Electronic address:  Cornwall@physics.ucla.edu\\
\newpage

\begin{center}
{\bf I.  INTRODUCTION}
\end{center}

\bigskip
The problem of the generation of cosmic magnetic fields is still open.
Some authors \cite{ka} argue that no primordial seed field is necessary, and
that Biermann-battery effects (non-vanishing $\grad n \times \grad T$
where $n$ is density and $T$ is temperature) act as a source for magnetic
fields at the time of structure formation, with the source due to shocks.
But many others search for primordial seed fields, associated with the
electroweak phase transition \cite{vach,c94,eo,bbm}, the QCD phase transition \cite{chol,qu}, or various other mechanisms, e.g. inflation \cite{tw}.  Efforts involving phase transitions are further subdivided according to their assumption of a first-order phase transition \cite{ho} or otherwise; a first-order transition produces bubbles and turbulence.

So far there has been little discussion of the generation of primordial magnetic helicity $H_M$, defined by what is otherwise known as a Chern-Simons term:
\begin{equation} H_M = \int d^3x \vec{A} \cdot \vec{B}.
\end{equation}
This P- and CP-odd function is important for several reasons:
\begin{itemize}
\item It is nearly conserved in the early universe (exactly so if the
conductivity is infinite).
\item It is not possible to have magnetic fields which are completely
homogeneous and carry helicity, so if there is primordial helicity there is
also some sort of spatial structure in the universe at early times.
\item If it were possible to measure a net helicity for the present universe
it would be another macroscopic manifestation of CP violation.\footnote{The most
important manifestation is
the very existence of the stars and galaxies at present abundances.  Much more speculatively, another could be the predominance of life forms of a single chirality.}
\item It is well-known to students of magnetohydrodynamics (MHD) that the
presence of magnetic helicity (or other parity-odd expectation values) can lead to unstable dynamo action \cite{can,kra,fplm,pfl}.
\end{itemize}

In the present paper we consider a scenario involving generation of
magnetic fields and helicity at the electroweak phase transition.  We can discuss the part involving just the generation of magnetic fields fairly precisely \cite{c94,cy,co94}; it involves a condensate of EW magnetic fields
in the magnetic $N=0$ Matsubara sector of the $SU(2)$ part of the EW 
gauge theory.  This sector of the theory is strongly-coupled and an $SU(2)$
magnetic condensate can be shown, without any approximation, to form \cite{co94}.
This condensate is characterized by a finite density of vortices (closed string-like objects which are very long, with a thickness inversely proportional to the dynamically-induced gauge-boson mass) which are randomly linked; the Chern-Simons number is a direct measure of this linkage \cite{co94,cy,mo,vf}.
(Thermal sphalerons may also be present, but their Boltzmann factor is likely to be small \cite{co89}.)  Like any other condensate, the vortex fields are sustained by a macroscopic number of phase-coherent W-bosons.  When the temperature falls substantially below the critical temperature $T_c$, the condensate loses phase coherence and is expressed as particles, plus Maxwell magnetic fields generated by the charged Ws.  These interact electromagnetically, retaining their original structure for a short while (because the W-boson mass does not change very much in the immediate neighborhood of the phase transition).  For these Maxwell fields the Chern-Simons number or helicity expresses a linkage between magnetic field lines, a linkage inherited from the W-condensate linkages.

This mechanism for magnetic field generation is somewhat different from Vachaspati's; it will be discussed
in more detail in Section II.  Although we do not even have the same scaling with the EW coupling $g$ that Vachaspati does, we find similar numerical values for the
magnetic field, of some $10^{23}$G at a magnetic scale length (transverse vortex size) $l_M\approx 2/g^2T\simeq 10^{-15}$ cm
appropriate for the magnetic $N=0$ sector.\footnote{Numerical values are taken from Refs. \cite{cy,nair,hkr}.  It is natural, in view of the numerical values, to quote the magnetic mass and condensate values in terms of $g^2$ rather than the corresponding EW fine-structure constant $\alpha_W = g^2/4\pi
\approx 1/30$.  Other quantities are more naturally expressed in terms of the fine structure constants themselves, so a mixture of these appears in various expressions.}

The question of helicity generation is far less well-understood, because
it is related to problems of formation and destruction of baryons (B) and
leptons (L) in the early universe.  We do not understand very well at all how B or L is generated in the early universe,
but we do know something about how B+L can be destroyed by sphalerons and other EW
effects at temperatures ranging from far above $T_c$ to just below it \cite{krs,c94,cy}.  These EW effects all involve the B+L anomaly,
so that changes in the number $N_{B,L}$ are necessarily accompanied by changes
in certain topological charges of the EW gauge theory.  But further precision in this picture is hard to come by, since we do not presently know whether the
present values of $N_B$ and $N_L$ can be generated by EW effects, possibly including effects of bubbles; whether it is due to primordial (and conserved) B-L generation; what the net balance between creation and destruction by EW effects is, and so on.  Nor do we
understand in detail how helicity generated above $T_c$ can survive to below
$T_c$ (but we will make some comments later in the spirit of the work of
Martin and Davis \cite{md} on the stability of magnetic fields generated at
phase transitions).  There are several mechanisms for translating the kind of EW helicity, or Chern-Simons terms, formed by B+L violation into magnetic helicity (of course, the usual Maxwell helicity does not occur in the
B,L anomalies, since these currents are vectorlike with respect to the
Maxwell field).  We will discuss these mechanisms briefly in Section III, and postpone further elaboration to work now in preparation.  For the present we will simply assume that the magnetic helicity surviving the EW phase transition is, give or take a couple of orders of magnitude:
\begin{equation} H_M = \alpha^{-1}(N_B+N_L) \simeq 10^{66} \; {\rm erg-cm}.
\end{equation}
The numerical value comes from taking $N_B \simeq 10^{80}$; a factor like
$\alpha^{-1}$, where $\alpha$ is the fine-structure constant, is reasonable from the form of the $B+L$ anomaly.

Another way of thinking about $H_M$ is to write its density as
\begin{equation} \vec{A} \cdot \vec{B} = \alpha^{-1}\epsilon T^3
\end{equation}
where the number density of any particle (or antiparticle) is $\sim T^3$ and $\epsilon$ is a small number presumably
related to, but not necessarily equal to, the B+L-entropy ratio (about $10^{-10}$ today\cite{kt}).  In view of the smallness of $\epsilon$ it may be
wondered whether the helicity we assume can have any important effects.  We argue that it does, on several grounds.  First, up to about the time of decoupling helicity on large spatial scales is very nearly conserved, and at about this time the primordial helicity is likely to be large compared to natural scales for helicity in the universe then.   In fact, although the helicity density is small compared to $T^3$ it is 30 or so orders of magnitude greater than one would expect from random fluctuations of the primordial magnetic fields in the absence of $B+L$-violating effects (just as the present $B+L$ number is about $10^{35}$ times the value expected from fluctuations alone).  Second, the presence of a primordial helicity can greatly affect the evolution of magnetic fields in the early universe, in particular by helping to generate an inverse cascade \cite{fplm,pfl,beo} in which the original magnetic fields at the time of the EW phase transition, which have a very short magnetic correlation length $l_M\simeq 10^{-15}$ cm
are transformed into fields on much longer scales.  (At the same time, there is a certain amount of destruction of the short-scale fields by magnetic viscosity;
the universe is not a perfect conductor.)

In Section IV we take up, in a simple approximation, the MHD issues concerning unstable dynamo growth driven by helicity in the early universe.  This approximation is similar in spirit to two-scale models or mean-field models (see, e.g., Ref. \cite{kra}), in which the effect of fields fluctuating on
small scales on long-scale fields is estimated by a process of spatially averaging quantities quadratic in the short-scale fields.  Our approximation is such that magnetic helicity $H_M$ is exactly conserved in the infinite conductivity approximation; finite-viscosity effects are trivially incorporated.
We find a quasi-linear instability in which helicity transfers itself and the magnetic fields from short to long scales.  There is an inverse-cascade instability for all lengths greater than a critical length scaling with $\epsilon^{-1}$, much larger than the magnetic length $l_M$ and much smaller than the Hubble size.  
It is perhaps surprising to find that although the maximum growth rate
associated with the inverse cascade of helicity is $O(\epsilon^2)$, this rate can be comparable to the expansion rate of the universe.  Part of this comes from the slowness of
the expansion of the universe compared to the natural EW rates (EW rates are
$\sim 10^{14}$ times the expansion rate), and part from the occurrence of inverse powers of fine-structure constants in the formulas.

Eventually the system approaches an equilibrium of the sort suggested in Refs.  \cite{kra,ta}, in which the final scale length is $\sim 1/\epsilon$.  This length is much bigger than $l_M$ but much smaller than the Hubble size.

If the primordial helicity is large compared to fluctuation effects, it is interesting to ask whether it could be measured.  We discuss this in Section V, and find it to be very doubtful, on two counts.  First, measuring helicity of whatever size would require not only very complete Faraday rotation data, but also data on the angular gradients of the magnetic field; these would be very hard to get, but in principle are available from scattering from polarized dust grains.  Second, the primordial helicity would dominate other effects up to about the time of structure formation, but after that time there are so many effects which process pre-existing magnetic fields that only very shaky conclusions could be drawn. 

\bigskip
\begin{center}
{\bf II. THE MAGNETIC EW CONDENSATE}
\end{center}

\bigskip

The general subject of the EW $SU(2)$ condensate has been considered in detail elsewhere \cite{c94,cy}.  At temperatures above $T_c$ the Higgs field has
vanishing VEV so the magnetic Ws are perturbatively massless in the $N=0$ Matsubara sector (in other Matsubara sectors the W-bosons either have an effective thermal mass $\sim T$ or, in the $N=0$ electric sector a perturbative mass $\sim gT$).  However, it is well-known that non-perturbative effects generate a magnetic mass $M_W \sim g^2T$ for the $SU(2)$ gauge bosons, or Ws (but no such mass is generated for the hypercharge bosons).  This mass is directly associated \cite{co94,lav} with a magnetic W-condensate
\begin{equation}
\langle \theta \rangle \equiv \frac{1}{4}\langle (G^a_{ij})^2 \rangle \sim g^6T^4,
\end{equation}
 and a negative free-energy density for the $N=0$ magnetic sector of value $-\langle \theta \rangle /3$.  Numerical estimates \cite{cy,ikps,kkrs} for the
free-energy density are in the neighborhood of $-(0.01-0.02)g^6T^4$.  Because the neutral W potential is related to the usual Maxwell potential by $A_i=\sin \theta_W W_i^3 + \cos \theta_W Y_i$, ($Y_i$ is the hypercharge potential) there is a condensate of Maxwell fields too, with (including a factor of $4\pi$ for
cgs units):
\begin{equation}
 \langle \vec{B}^2 \rangle
\simeq 0.5 e^2g^6T^4 
\end{equation}
and an RMS Maxwell field strength of about $10^{23}$G at
$T\simeq T_c$.  As already mentioned, 
the correlation length of this field above the EW phase transition is
$l_M \simeq 2/g^2T \simeq10^{-15}$ cm at $T\simeq T_c$.  (For comparison, the EW Hubble scale is about 1 cm.)  

This field strength at the phase transition is numerically comparable to Vachaspati's estimate \cite{vach}, but the
scalings seem to be different.  This is because he takes the W-mass to scale
like $gT$ instead of $g^2T$. Both the present work and Vachaspati agree that
$B\sim M_W/g$, but we use the usual magnetic-mass scaling for $M_W$.  The numerical values
agree fairly well because $g\approx 0.65$ is not small compared to unity.

The spatial structure of this field is of interest.  It could have finite-$T$
sphalerons as a component, but as mentioned earlier \cite{co89} the mass of the
magnetic sphaleron is so large that its Boltzmann factor is likely to be rather small, and we will ignore sphalerons.  The dominant component is a gas of closed vortices of thickness $M_W^{-1}$ whose entropy dominates the internal energy, so the vortices or strings like to be very long.  They can link with each other and with themselves\footnote{Self-linking is also known as twisting and writhing; see  \cite{c94} for a discussion and references.}, and also with
Wilson loops; this latter is the mechanism for confinement both in three and
four dimensions (in $d=4$ the vortices are closed two-surfaces) \cite{co79}, and the
string tension is an outgrowth of fluctuations of the linking numbers of the
vacuum condensate with the Wilson loop.  The sum of all linking numbers of the vacuum (plus the sphaleron number, if any) is proportional to the Chern-Simons number of the thermal vacuum \cite{c94,cy}.
This will be zero in the absence of such parity-violating effects as
B+L violation; in its presence, the vacuum will have a net linking number.
It is this linking number that we argue will be (in part) preserved as a
contribution to Maxwell helicity after the EW phase transition is complete.

We postpone a complete discussion of the transition dynamics to a later work.  The general picture is that as the W-condensate dissolves (and the Higgs condensate forms), the phase-coherent Ws forming the condensate lose this coherence and become particles.  We therefore need a picture of the condensate in particle language.  Think first of a liquid-helium condensate:  The atoms forming it are (in the center-of-momentum frame) strictly at zero momentum, even though the condensate coexists with other particles of finite momentum.  The EW condensate is a little different because of the spatial structure of the vortices, which requires the Ws to have finite momentum, but smaller than the thermal momentum $\sim T$.  In plasma language the vortex somewhat resembles a
$\theta$-pinch, with pressure gradients balancing $\vec{J}\times \vec{B}$ forces.  By looking at the vortex solutions (see the work of Cornwall in Ref. \cite {co79}) one reads off the scalings for $SU(2)$ magnetic field $B$ and condensate current density $J$ and then infers the condensate pressure $p$:
\begin{equation}
 B\sim M_W^2/g;\;\;J\sim M_W^3/g;\;\;p\sim g^6T^4.  
\end{equation}
The current is also $ng$, where $n$ is the density of particles of one sign of charge, and the current velocity is of order unity.  With $M_W\sim g^2T$, one finds that $n\sim g^4T^3$, so that a relatively small fraction of the particles in the thermal bath are tied up in the condensate.  The particle energy in the condensate is $nM_W\sim g^6T^4$, like the magnetic energy in Eq. (4).  The Larmor radius of a particle scales like $M_W/gB\simeq 1/g^2T=l_M$.  The condensate particles and fields are in equilibrium above the phase transition, and depart from it relatively little during and just after the phase transition, in part because the W mass does not change very much during the transition.  Of course, the photon mass does change drastically\footnote{In fact, there is a sense (see Section 4) in which the photon mass becomes very slightly tachyonic, of $O(\epsilon)$ .} and new magnetic fields are generated, but their currents have for a while their original correlation length $l_M$.

\bigskip
 
\begin{center}
{\bf III.  B+L-VIOLATING GENERATION OF MAGNETIC HELICITY}
\end{center}

\bigskip

We are ignorant about the actual mechanism of B+L generation in the universe, although many candidates abound.  The best we can do at present is to construct a scenario which is not obviously wrong; this scenario will not have precise
numbers associated with the B+L violation effects. 
Imagine, then, that EW effects are well-described by the Standard Model, with no supersymmetry or other non-standard mechanisms being important.   We will assume that the EW phase transition is second order (meaning a Higgs 
mass larger than about 80 GeV).  In this case, there is no B+L production
and no turbulence associated with the EW phase transition.  The EW effects are
limited to dissipation of some part of whatever B+L has been previously generated by some unknown mechanism.  To avoid fine tuning, we assume that the original amount of B+L is comparable to what exists after the phase transition, or otherwise there is a cancellation between a large production rate and a large dissipation rate, leaving a much smaller net amount.  By this assumption the amount of B+L which has been dissipated by EW effects is comparable to what exists today.  

We can roughly model these dissipation effects and their consequent generation of topological charge by introducing a chemical potential associated with the
conserved (but gauge-variant) current which is the sum of the B+L current and a topological current:
\begin{equation} K_{\mu} = J_{\mu} + \frac{n_f}{8\pi^2}\epsilon_{\mu\nu\alpha\beta}\{Tr[g^2A_{\nu}\partial_{\alpha}
A_{\beta} - \frac{g}{3}A_{\nu}A_{\alpha}A_{\beta}] + \frac{1}{2}g'^2Y_{\nu}
\partial_{\alpha}Y_{\beta}\}
\end{equation}
in standard notation.  The grand partition function
\begin{equation}
Z=Tr e^{-\beta (H-\mu Q)},
\end{equation}
where $Q=\int d^3x K_0$ is the conserved charge, reinstates invariance under large gauge transformations
by summing over all possible values of $N_{B+L}$.  The chemical potential
$\beta\mu$ is of order $n_{B+L}T^{-3}$, that is, $O(\epsilon)$.  As far as the
$N=0$ magnetic Matsubara sector of the $SU(2)$ gauge fields is concerned, this
chemical potential introduces a new gauge-coupling term $\sim \epsilon$ which is CP-odd.

There is, of course, no term in the B+L charge which directly involves the Maxwell
helicity $\int d^3x \vec{A} \cdot\vec{B}$, because of the vectorlike nature of
electromegnetism.  But there are small terms in $\mu Q$ linearly coupled to $\vec{A}$,
which can be treated as perturbations on the underlying $SU(2)$ condensate, in which electromagnetism participates.  
Since there is no hypercharge condensate\footnote{At least, as driven by standard model fields; the $Y$ boson remains massless
at $T\geq T_c$.  Other sources of a $Y$ condensate have been envisaged; see
Joyce and Shaposhnikov \cite{js}.} we can just set the hypercharge potential to
zero, so that one component (conventionally the third) of $W_i$ is $\sin \theta_W A_i$.  This is, at $T\geq T_c$, a massive field with vortex solutions
like the Nielsen-Olesen vortices; these solutions have been discussed in
several places \cite{co79,c94,cy,co96}.  Let us write the effective CP-odd
action for the Maxwell field as
\begin{equation}
\mu\int d^3x \vec{P}\cdot\vec{A}.
\end{equation}    
Given that the underlying $SU(2)$ condensate is one of strings, it turns out
that $\vec{P}$ is effectively a magnetic field of the type
\begin{equation}
P_i = \sum q\oint dz_i M^2\Delta_M (\vec{x}-\vec{z})  
\end{equation}
where $M$ (previously denoted $M_W$) is the magnetic mass, $\Delta_M$ is the massive Euclidean propagator for mass $M$, the integral runs over a closed string, the sum is over the
collective coordinates of the strings in the condensate, and $q$ summarizes
various constants and group matrices of no concern to us now.  In a gauge
where $\grad \cdot \vec{A}=0$, the form of $\vec{A}$ to first order in
$\epsilon$ is:
\begin{equation}
A_i(\vec{x})=c\sum \epsilon_{ijk}\partial_j\oint dz_k\{\Delta_M-\Delta_0\}(\vec{x}
-\vec{z})+(\nabla^2+M^2)^{-1}\mu P_i
\end{equation}
($\Delta_0$ is the massless propagator).  Again the sum is over the strings
of the condensate, and $c$ is a collection of constants of no interest now.
This expression for $\vec{A}$ is reminiscent of a similar expression \cite{co96} for a vortex in $d=3$ Yang-Mills theory with a Chern-Simons term added.  Even if such a vortex string is straight one finds that it has a link number, coming from the twist of the field lines.  Similarly, one readily calculates the magnetic helicity $\int d^3x \vec{A}\cdot\vec{B}$ from Eq. (11)
to find that there is an $O(\epsilon)$ term from the product of the
$\epsilon_{ijk}$ piece and the $P_i$ piece, which, in its structure of
string integrals, resembles a sum over regulated Gauss linking numbers.

The calculation we have sketched out here relates the net $SU(2)$ helicity, coming from B+L violation, to the currents driving the Maxwell magnetic field.
It holds for $T>T_c$.  The next question to ask is how much of the magnetic
helicity can survive the EW phase transition.  If the problem is restricted to asking how much of the magnetic field, regardless of its helicity, survives, an answer has already been given by Martin and Davis \cite{md}.  Their answer is that the field does survive largely intact.  Although we will not discuss it in detail, we believe that most of the helicity also survives.  One approach is similar to that of Martin and Davis:  Just below the phase transition, when the
Higgs field $\phi$ has a VEV (denoted $v$) one can use \cite{vach} 't Hooft's \cite{th}
expression for the Maxwell field strength involving projecting out the Abelian part with the Higgs field.  This can be written:
\begin{equation} F_{ij}=\partial_iA_j -\frac{4i}{gv^2}\sin \theta_W
(\partial_i\phi)^{\dagger}\partial_j\phi - (i\leftrightarrow j).
\end{equation}
where the electromagnetic vector potential $A_i$ is defined in terms of the
unit Higgs vector as:
\begin{equation} 
A_i=\sin \theta_W n^aW^a_i;\;\;n^a=\phi^{\dagger}\sigma^a
\phi/v
\end{equation}
(There is also an electric field which we will not consider here.)
The phase of the Higgs field near the transition temperature inherits the topological information on the linkage of the string condensate.  Martin and
Davis consider thermal fluctuations at a second-order phase transition and 
conclude that these do not wipe out the magnetic field.  The same conclusion ought to hold for the helicity as well.

Another purely classical argument has been given by Taylor \cite{ta} for
Tokamak plasmas, which have macroscopically linked magnetic fields.  He begins with a non-equilibrium plasma, of the sort
that might be produced at a phase transition, and points out that for closed field lines in the ideal MHD case, the magnetic helicity ought to be preserved separately for every closed field line (or surface).  He then argues that as the plasma relaxes to equilibrium through effects due to large but finite conductivity and field lines reconnect, the helicities on the various field lines will be homogenized, but the sum (i.e., volume integral) of all helicities should be preserved, because reconnection---which homogenizes helicity---does not make large changes in the fields and potentials themselves. 

Given that some helicity of order $\epsilon$ survives, how does it affect the evolution of magnetic fields after the EW transition?  We discuss this next.

\bigskip

\begin{center}
{\bf IV.  MAGNETOHYDRODYNAMICS AND MAGNETIC HELICITY}
\end{center}

\bigskip

Although it is known \cite{kra,can,fplm,pfl} that helicity is an important driver of unstable MHD dynamo action, the problem we face is somewhat different from the standard MHD dynamo problem because general relativity must be taken into account \cite{beo} and because we consider an initial-value problem rather than the often-considered problems of steadily-driven helicity.   

Let us consider some typical length scales, beginning with the Larmor radii.  In the $SU(2)$ condensate above the phase transition, the coherent Larmor radius $R_L$, or the radius a charged particle in the condensate would have if circulating around a single vortex, is about
\begin{equation}
T\geq T_c:\;\;R_L \approx g^2T/gB \sim 1/g^2T;
\end{equation}
a non-condensate particle would have a Larmor radius bigger by only a few, a
factor of $1/g^2$, modified possibly for the fact that it is circulating around a couple of vortices so the RMS B-field is a little smaller than the coherent field.  In the plasma after the phase transition some factors of $g$ change to factors of $e$ (compare Eqs. (4) and (5)), and the former condensate particles have a Larmor radius
\begin{equation}
T<T_c:\;\;R_L \approx 1/e^2T,
\end{equation}
a few times larger than before the phase transition.
These Larmor radii are bigger than Coulomb collision lengths, which scale like
$1/\alpha^2T$ (for quarks the QCD length is about $1/\alpha_S^2T$; these could be unmagnetized, but they play no role in the condensate formation anyhow).
In what follows, we will assume that the Larmor radius of Eq. (15) is appropriate for the inital scale lengths of Maxwell fields just after the phase transition.

The helicity-driven dynamo picture is that these small-scale fields drive an instability of fields on larger scales.  As is usual in such cases \cite{kra,fplm,pfl}, we will average quantities quadratic in fields varying on small scales to get driving terms on the large scales.
        
   After the modifications for general relativity our approach is similar to that of
Ref. \cite{kra} for so-called $\alpha$ dynamos, but with some significant differences.   Brandenburg, Enqvist, and Olesen \cite{beo} have shown that the ideal MHD equations (no kinetic or magnetic viscosity) have simple scaling properties in a flat Robertson-Walker metric
\begin{equation}
ds^2=dt^2-R^2(t)d\vec{x}^2
\end{equation}
which reduces them to the flat-space MHD equations provided that cosmic time
$t$ is replaced by conformal time  $\tilde{t}$:
\begin{equation} \tilde{t}=\int dt/R(t) = 2(t_{EW}t)^{1/2}
\end{equation} 
where we normalize so that at EW time $t_{EW}\approx 10^{-11}$ sec the scale
factor $R(t)$ is unity.  The required ideal-MHD scalings are:
\begin{equation}B=\tilde{B}/R^2,\;\;J=\tilde{J}/R^3,\;\;\rho=\tilde{\rho}
/R^4,\;\;v=\tilde{v}.
\end{equation}
Here $B,J,\rho,v$ are respectively the magnetic field, the current, the MHD energy density, and the bulk velocity.  Other scalings can be inferred from these and the statement that only time is changed, not spatial gradients.
The meaning of these scalings is that the variables with a tilde do not change just because the scale factor $R$ is changing.
The scaling of magnetic field reflects the usual dilution with flux conserved.

 We need two additional scalings, since we will consider non-ideal MHD.  These scalings are for the conductivity $\sigma$ and the collision (or correlation) time $\tau$.  In the case of Coulomb collisions it is well-known that
\begin{equation} 
\sigma \approx T/\alpha =\tilde{\sigma}/R;\;\;\tau_c\approx 
(\alpha^2T)^{-1} = \tilde{\tau}_c R.
\end{equation}
The magnetic viscosity $\nu_M$ is $1/\sigma$ and scales like $R$; so does the
kinetic viscosity (and also a kind of Alfv\'en viscosity defined later).

If one writes out the general-relativistic MHD equations in terms of the
original variables one finds that they are the same as the flat-space MHD
equations with the tilde variables.  Before we write these equations down we make two points:  First, the bulk velocity $\vec{v}$ is expected to be non-relativistic, even if the phase transition is first-order.  Of course, the thermal velocities of the (non-condensate) plasma are essentially the speed of light, so the energy density $\rho$ is quite relativistic.  Second, we will assume a second-order transition, which means that there should be no turbulence in the matter variables and that, because of inflation, the scale lengths for these variables are quite large, at least initially.\footnote{But the scale lengths for the current $J$ are the same as for the magnetic field and helicity, that is, of $O(1/e^2T)$.  There is no reason for the scale length for other matter variables to be this short.}  As the instability we identify below grows, the matter variables at later times will also become turbulent, which can have important effects \cite{pfl} not considered here. 

To avoid notational complexity we drop the tildes on the scaled variables except for that denoting conformal time.  We also drop all gradients referring to matter variables, including the pressure and kinematic viscosity terms as well as the non-linear velocity term.\footnote{It is only the pressure gradient that is dropped; the pressure contributes to the overall energy density.}  For the moment we will keep magnetic viscosity.  Then the matter equation is (using $\vec{J}=\grad\times\vec{B}$)
\begin{equation}
(4/3)\rho \frac{\partial \vec{v}}{\partial \tilde{t}} =
-\vec{B}\times (\grad \times \vec{B}).
\end{equation}
Using the electric-field relation
\begin{equation}
\vec{E} = -\vec{v}\times\vec{B} + \vec{J}/\sigma
\end{equation}
we can strip off a curl operation from the $\vec{B}$ equation to get:
\begin{equation}
\frac{\partial \vec{A}}{\partial \tilde{t}} = \vec{v}\times\vec{B}
-\frac{1}{\sigma}\grad\times\vec{B}+\grad\chi.
\end{equation}
The $\grad\chi$ term has no physical effect and can be dropped.  

Magnetic viscosity is due to the $1/\sigma$ term, and is dominated by Coulomb scattering (particle viscosity is dominated by quark scattering).  A numerical estimate at
the EW phase transition gives
\begin{equation}
\nu_M \equiv 1/\sigma \approx 10^{-8}-10^{-9} {\rm cm}^2/{\rm sec}.
\end{equation}  The corresponding magnetic Reynolds number $lv/\nu_M$
ranges from about $10^5$ at the correlation length scale $l\approx 1/\alpha_WT$ to $10^{19}$ at the Hubble scale $l=H^{-1}$, assuming that the velocity is the speed of light (actually it is somewhat smaller).  We will identify below a sort of Alfv\'en magnetic viscosity which is somewhat larger than the collisional
viscosity, and therefore drop the $1/\sigma$ term in Eq. (22).

The next step in principle is to solve for the velocity from Eq. (20).  Of course this cannot be done analytically, and we make a common approximation (see,  e.g., Ref. \cite{pfl,gd}) and invert the $\tilde{t}$ integration in Eq. (20) by
multiplying by a correlation time $\tau$. It is reasonable to choose $\tau$ to be the Coulomb time  $\tau_c\approx 1/\alpha^2T$ (most particles do not change their velocity appreciably in one Larmor period). Assuming that the initial velocity vanishes:
\begin{equation}
\vec{v} = -\frac{3\tau}{4\rho}\vec{B}\times(\grad\times\vec{B}).
\end{equation}
Then the equation for $\vec{A}$ is:
\begin{equation}
\frac{\partial \vec{A}}{\partial \tilde{t}}=\frac{3\tau_c}{4\rho}
\{\vec{B}[\vec{B}\cdot(\grad\times\vec{B})]-(\grad\times\vec{B})B^2\}.
\end{equation}

This highly-nonlinear equation is intractable as it stands, so we quasi-linearize it by replacing two of the three factors involving the magnetic field on right-hand side by averages  (in the spirit of the mean-field dynamo \cite{kra} in which averages are made over short spatial scales to find their effect on long scales).  We assume, by isotropy, that all vectorial quantities, in particular $\vec{B}$, have zero average, so the only possibility is to replace Eq. (25) by:
\begin{equation}
\frac{\partial \vec{A}}{\partial \tilde{t}}=\frac{3\tau_c}{4\rho}
\{\vec{B}\langle \vec{B}\cdot(\grad\times\vec{B}) \rangle -(\grad\times\vec{B})\langle B^2 \rangle\}.
\end{equation}
This is of the canonical form of an $\alpha$-dynamo with turbulent viscosity\cite{kra}:
\begin{equation}
\frac{\partial \vec{A}}{\partial \tilde{t}}=\alpha_M\vec{B}-\beta_M(\grad\times\vec{B}).
\end{equation}

So far we have not been specific about the nature of these averages.  To do that, we note that once
having dropped the magnetic viscosity, it should be that the magnetic helicity is conserved.  Compute the rate of change of helicity in a large volume $V$:
\begin{eqnarray}
\dot{H}_M & = & 2\int_V d^3x  \frac{\partial \vec{A}}{\partial \tilde{t}}\cdot
\vec{B}\nonumber\\
& =& 2\int_V d^3x \{\alpha_M B^2-\beta_M [\vec{B}\cdot (\grad\times\vec{B})]\}.
\end{eqnarray}
Clearly if one defines averages so that
\begin{equation}\alpha_M=\frac{3\tau_c}{4V\rho}\int_V d^3x \vec{B}\cdot(\grad\times
\vec{B}),\;\;\beta_M=\frac{3\tau_c}{4V\rho}\int_V d^3x B^2
\end{equation}
then helicity is conserved.

One may now go back and add the collisional magnetic viscosity $\nu_M$ to
$\beta_M$ in Eq. (27), an addition which will lead to destruction of helicity on
short scales.  We will not consider the collisional magnetic viscosity further.

Although $\alpha_M$ is not the magnetic helicity $H_M=\int \vec{A}\cdot\vec{B}$
defined earlier, it is qualitatively equivalent to it, for the initial magnetic field configuration from which we start, which is dominated by a single scale length $l_M$ or $R_L$.  We will estimate the coefficients $\alpha_M,\;\beta_M$ just at the EW transition point.  At later times these values will change, not only because of $R$-scaling but also because of MHD effects.
Let us estimate $\alpha_M$ in terms of $H_M$.  To find $\langle \vec{B}\cdot (\grad \times \vec{B})\rangle$ we divide $H_M$ by the square of the Larmor  length $R_L\approx 1/e^2T$.
This yields:
\begin{equation}
\alpha_M \simeq \frac{\tau_c \vec{A}\cdot \vec{B}}{\rho R_L^2} \simeq
0.01\frac{\epsilon e^4}{\alpha^3}\approx 3 \times 10^{12}\epsilon \;{\rm cm/sec}.
\end{equation}
(The numerical factors in Eqs. (30) and (31) below summarize various factors involved in the densities, etc; we are using units such that $e\simeq 0.3,\;g\simeq 0.65$.)
For $\beta_M$:
\begin{equation}
\beta_M \simeq \frac{\tau_c}{\rho} \langle B^2 \rangle \simeq \frac{0.002e^2g^4}
{\alpha^2T} \approx 10^{-6} \;{\rm cm}^2/{\rm sec},
\end{equation}
considerably larger than the collisional magnetic viscosity of Eq. (23).
The ratio $\beta_M/\alpha_M$ defines a critical length, which we call $l_H$,
separating the larger scales to which helicity is transferred from the smaller scale from which it comes.  This is approximately
$3\times 10^{-18}\epsilon^{-1}$ cm.  So $l_H$ is much larger than any correlation length but much smaller than the Hubble size.  

Note that $\beta_M$ can be written as $\tau_c V_A^2$ in terms of a (nominal)
Alfv\'en velocity $V_A \sim eg^2$.  This suggests that the $\beta_M$-effect can be interpreted in terms of Alfv\'en waves carrying off helicity from small scales to large.  One can, in fact, easily check that Alfv\'en waves do carry helicity, but it is not clear that the Alfv\'en wave concept is suited to a plasma of the type found in the early universe.  

Next we study the quasi-linear instability resulting from equation Eq. (27), replacing $\alpha_M,\;\beta_M$ by constants.  In fact, these quantities change, so the approximation of constant values can only hold for a short time.  
There have been numerous studies of $\alpha$-dynamos, described by equations like Eq. (27).  We give a treatment
slightly different from any of which we know.  Begin with the standard \cite{kra} decomposition of the magnetic field into poloidal ($P$) and toroidal ($T$)
components:
\begin{equation} 
\vec{B} = \vec{L}T + \grad \times (\vec{L}P)
\end{equation}
with the magnetic potential in a natural gauge:
\begin{equation}
\vec{A} = -\vec{r}T + \vec{L}P.
\end{equation}  In these equations, 
\begin{equation}
\vec{L} \equiv \vec{r}\times \grad
\end{equation}
is the anti-Hermitean generator of angular momentum.
The helicity is given by:
\begin{equation}
H_M = \int d^3x \vec{A}\cdot\vec{B} = -2\int d^3x T L^2P
\end{equation}
and expresses the linkage between toroidal and poloidal field lines, both of
which must be present for there to be helicity.  (This fact is related to the
possibility of measuring cosmic helicity, which we discuss in Sec. V.)  
The equations for $T,P$ are easily derived:
\begin{equation}
\dot{T} = -\alpha_M \nabla^2P + \beta_M \nabla^2T
\end{equation}
\begin{equation}
\dot{P} = \alpha_MT + \beta_M \nabla^2 P
\end{equation}
These are straightforwardly solved by Fourier transforms (indicated by a hat):
\begin{equation}
T(\vec{x},\tilde{t})=\frac{1}{2(2\pi )^3}\int d^3k [\hat{T}_+
e^{\lambda_+\tilde{t}}+\hat{T}_-e^{\lambda_-\tilde{t}}]
\end{equation}
\begin{equation}
P(\vec{x},\tilde{t})=\frac{1}{2(2\pi )^3}\int \frac{d^3k}{k} [\hat{T}_+
e^{\lambda_+\tilde{t}}-\hat{T}_-e^{\lambda_-\tilde{t}}]
\end{equation}
where
\begin{equation} \hat{T}_{\pm}=\hat{T}\pm \hat{P}|_{\tilde{t}=0},\;\; \lambda_{\pm} = \pm k\alpha_M-k^2\beta_M.
\end{equation}
Whatever the sign of $\alpha_M$ there is growth (unless the corresponding
$\hat{T}$ identically vanishes); to be definite assume $\alpha_M >0$.
The maximum growth rate occurs for $k\approx l_H^{-1}$, and is given by
\begin{equation}
\Lambda \equiv Max\; \lambda_+ = \frac{\alpha_M^2}{4\beta_M} \approx 3\times  10^{30}\epsilon^2 \;{\rm sec}^{-1}.
\end{equation}
Since the expansion rate of the universe $H$ is about $10^{11}\;{\rm sec}^{-1}$, there will be appreciable dynamo growth for $\epsilon\geq 2\times 10^{-10}$, a not 
unreasonable number.\footnote{Recall that just after EW times the fractional baryon number is about $3\times 10^{-8}$, rather larger than it is now, because more photons have been produced at annihilation \cite{kt}.}

As Ref. \cite{kra} has pointed out, there are also equilibrium solutions to the equations (36, 37).  Naturally, their scale length is about $l_H$.  The equilibrium solution for $P$ or $T$ is {\em tachyonic}:
\begin{equation} \nabla^2 P + (l_H)^{-2}P = 0.
\end{equation}
Taylor \cite{ta} has given another understanding of this tachyonic equilibrium
equation, by introducing a real Lagrange multiplier corresponding to the conserved helicity $H_M$, which is a Chern-Simons term with real coefficient.  It is well-known that such a system is tachyonic.  However, no real problems are encountered; the solutions to Eq. (42), involving Bessel functions (for a given angular momentum), can be terminated smoothly matching on to multipolar magnetic fields in the region where the helicity vanishes \cite{kra}.

It is not possible for us to go to large times (say, larger than the expansion time of the universe) by analytical means, because the coefficients $\alpha_M,\;\beta_M$
are changing as the magnetic scale lengths change, and as magnetic energy is exchanged with fluid energy.    We hope to carry out elsewhere a detailed numerical simulation of the
MHD processes so crudely described here.  This simulation, going to later times, would reveal the influence of a number of effects we have omitted, including the growth of kinetic helicity in the MHD velocity field, as measured by $\int \rho \vec{v}\cdot (\grad\times\vec{v})$.  This should grow from its initial value of zero to approximate equipartition with the magnetic helicity as measured by 
$\int \vec{B}\cdot(\grad\times\vec{B})$, at which point one encounters the so-called Alfv\'en effect of Ref. \cite{pfl}:  These two helicities tend to cancel out on short spatial scales, leaving still a residual inverse-cascade instability.

\bigskip

\begin{center}  
{\bf V.  CAN WE DETERMINE THE EXISTENCE OF PRIMORDIAL HELICITY EXPERIMENTALLY?}
\end{center}

\bigskip

We remark on our assumed scale for magnetic helicity (see (2)) in comparison with the value expected from fluctuations at EW time, and on the possibility of measuring helicity today.

Supposes that at EW time the net helicity of the universe were given only by fluctuations of random-sign helicity in cubes the size of the magnetic correlation length $l_M$.  The maximum helicity in any cube scales like
$\langle B^2\rangle l_M^4$, so the RMS fluctuation value is found by multiplying by the square root of $N\approx V/l_M^3$, where $V$ is the volume of the universe at EW time.  Taking $\langle B^2\rangle$ from (5) one finds, with
$V\approx 10^{42}\;{\rm cm}^3$, that the helicity due to fluctuations is about
$10^{30}$ erg-cm, far smaller than we estimated in equation (2).  This is, of course, about the same as for the actual number of baryons today compared to the
fluctuation value of about $10^{45}$.  

If we could measure the EW-time helicity it would be easily possible to prove or disprove our hypothesis that there is a primordial helicity proportional to the baryon number.  Unfortunately we cannot, and the enormous processing of magnetic fields that must have taken place after matter dominance and structure formation would in any case greatly obscure any interpretation of a measurement of helicity today (that is, since structure formation).  But it may still make some sense to ask whether we could in principle measure the present magnetic helicity of the universe.

We have seen from Eq. (35) that to measure helicity requires simultaneous determination of both $T$ and $P$ in the decomposition Eq. (32) of the magnetic field.  Faraday rotation can, in principle at least, recover the component
$\vec{r}\cdot\vec{B}$, equivalent to $P$ (given enough sight lines to polarized sources and independent measurements of the electron density).  However, a separate measurement is needed, which must amount to the extraction of
$\vec{L}\cdot\vec{B}$, equivalent to $T$.  Again in principle, this could be measured from polarization of starlight by scattering from dust grains polarized in the cosmic magnetic field, given enough separate dust clouds, and a hypothesis of general isotropy of the magnetic field.

Finally, one might think of measuring a net circular polarization in the cosmic microwave background, but if this is only $O(\epsilon)$ this too is probably impossible.

Even if we could do this, would the primordial helicity of Eq. (2) be large compared to fluctuation-driven helicity today, as it was at EW times?  The answer is no, because even for a single galaxy with $B\sim 10^{-6}G$, an in-plane correlation scale of 15 kpc and a vertical correlation scale of 0.5 kpc the maximum helicity is $\sim 10^{77}$ erg-cm, far greater than a primordial value.  This very large value on the primordial scale is analogous to the very large value of the fields and coherence lengths compared to those of primordial fields (scaled to the time of structure formation) with or without helicity, and simply reflects the well-known fact that today's fields grew exponentially during the time since structure formation, by dynamo action.  (It is, of course, also possible that
primordial fields and helicity have nothing to do with today's cosmic magnetic fields.),
Still, measuring the helicity of today is interesting in principle, since it might very well be larger than one would expect from fluctuations alone, even though there has been so much processing of magnetic fields since the beginning of structure formation.
  
\newpage

\begin{center}
{\bf ACKNOWLEDGMENTS}
\end{center}

\bigskip

This work was begun at the Aspen Center for Physics in the summer of 1996, and we thank the Center for hospitality.  We also thank B. Chandran, G. B. Field, T. Vachaspati,
R. Rosner, and K. Enqvist for helpful discussions at Aspen, and S. Cowley and
Chandran for discussions at UCLA.

This work was supported in part by the National Science Foundation under 
grant PHY 9531023.

\end{document}